\documentstyle[twoside,fleqn,espcrc2,epsf]{article}

\title{
\vspace{-4.2cm}
\begin{flushright}
{\normalsize
hep-lat/9909049\\
UTCCP-P-71\\
UTHEP-409\\
\vspace{-4pt}
September 1999}
\end{flushright}
\vspace{1.5cm}
Quenched QCD with domain-wall fermions on coarse lattices
\thanks{Talk presented by Y.~Aoki at Lattice 99, Pisa, Italy.}}

\author{CP-PACS Collaboration : 
  A.~Ali Khan\rlap, \address{Center for Computational Physics,
    University of Tsukuba, Tsukuba, Ibaraki 305-8577, Japan}
  S.~Aoki\rlap,\address{Institute of Physics,
    University of Tsukuba, Tsukuba, Ibaraki 305-8571, Japan}
  Y.~Aoki\rlap,$^{\rm a,b}$
  R.~Burkhalter\rlap,$^{\rm a,b}$
  S.~Ejiri\rlap,$^{\rm a}$
  M.~Fukugita\rlap,\address{Institute for Cosmic Ray Research,
    University of Tokyo, Tanashi, Tokyo 188-8502, Japan}
  S.~Hashimoto\rlap,\address{High Energy Accelerator Research Organization
    (KEK), Tsukuba, Ibaraki 305-0801, Japan}
  N.~Ishizuka\rlap,$^{\rm a,b}$
  Y.~Iwasaki\rlap,$^{\rm a,b}$
  T.~Izubuchi\rlap,$^{\rm b}$
  K.~Kanaya\rlap,$^{\rm a,b}$
  T.~Kaneko\rlap,$^{\rm a}$
  Y.~Kuramashi\rlap,$^{\rm d}$
  T.~Manke\rlap,$^{\rm a}$
  K.~Nagai\rlap,$^{\rm a}$
  M.~Okawa\rlap,$^{\rm d}$
  H.P.~Shanahan\rlap,\address{DAMTP, University of Cambridge, 
    Cambridge, CB3 9EW, England, UK}
  Y.~Taniguchi\rlap,$^{\rm b}$
  A.~Ukawa\rlap,$^{\rm a,b}$ and
  T.~Yoshi\'e$^{\rm a,b}$
  }

\begin{document}

\begin{abstract}
We investigate the existence of chiral zero modes at  $a^{-1} \simeq$ 1 GeV 
in quenched domain-wall QCD.
Simulations are carried out for the plaquette and an RG-improved gauge actions
on a $12^3\times 24\times N_s$ lattice with $N_s=10-50$.
We find that the pion mass in the chiral limit remains 
non-vanishing as $N_s\to\infty$ for both gauge actions.
Possible origins of this non-vanishing pion mass are discussed.
\end{abstract}

\maketitle

\section{INTRODUCTION}

The domain-wall fermion formulation of QCD (DWQCD)~\cite{Kaplan,FS} 
is expected to
realize exact chiral symmetry on the lattice without species doubling 
at finite lattice spacing. 
This represents an appealing possibility, 
particularly for investigations of problems 
such as weak matrix elements sensitive to chiral 
symmetry~\cite{BlumSoni}.  Therefore, 
a number of studies have been made~\cite{BlumRev}.

Simulations in DWQCD, however, requires considerable computing power.
Even if the size of the extra dimension $N_s$ can be taken as small, 
i.e.\ $N_s=O(10)$, 
lattice spacings much finer than $a^{-1}\approx  2$ GeV will be difficult to 
simulate even in quenched QCD.  Hence simulations on coarse lattices down to 
$a^{-1}\approx  1$~GeV will be needed for reliable continuum extrapolations.

A first step in DWQCD at such a strong coupling 
will be to ensure the existence of chiral zero modes.  
Here we report our preliminary results on this problem. 
Our study is made for the plaquette action, and also for an RG-improved 
action~\cite{RG} to examine the effects of reduced discretization errors. 

\section{PARAMETERS}

We employ the fermion action identical to that in Ref.~\cite{FS},  
with the domain wall height $M$ and  bare quark mass $m_f$.
The gauge coupling is chosen to be $\beta=5.65$ for the plaquette action 
and $\beta=2.2$ for the RG action; these values correspond to 
the scale $a^{-1}\approx 1$ GeV determined from the string tension. 
Simulations are made on an $12^3\times 24\times N_s$ lattice with 
$N_s=10, 20, 30$ and 50.  

In the free fermion case the chiral zero mode exists over the range $0<M<2$. 
Since this range will be shifted in the interacting theory, we employ 
$M=1.3, 1.7, 2.1$ and 2.5.  For each of these values of $M$, we take 
$m_f=0.1, 0.05$ and 0.03 and calculate the pion mass for both degenerate and 
non-degenerate quark and antiquark pairs. 
For each parameter point we have typically $20$ configurations.

\section{RESULTS}

\begin{figure}[t]
  \vspace{-16pt}
  \begin{center}
    \leavevmode
  \epsfxsize=7cm \epsfbox{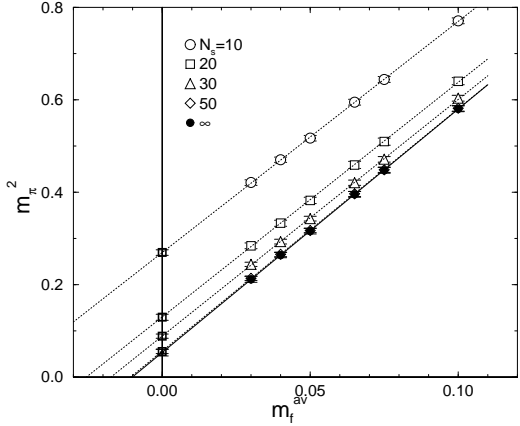}
  \end{center}
\vspace{-32pt}
  \caption{Pion mass squared as a function of $m_f^{av}$ 
    at $M=1.7$ for the plaquette action.
    Solid circles show the $N_s\rightarrow \infty$ limit of open symbols
    for each $m_f$.
    Lines show the linear fits.
    }
  \label{fig:pi2-mfNs}
  \vspace{-20pt}
\end{figure}

\begin{figure}[t]
  \vspace{-6pt}
  \epsfxsize=7.6cm \epsfbox{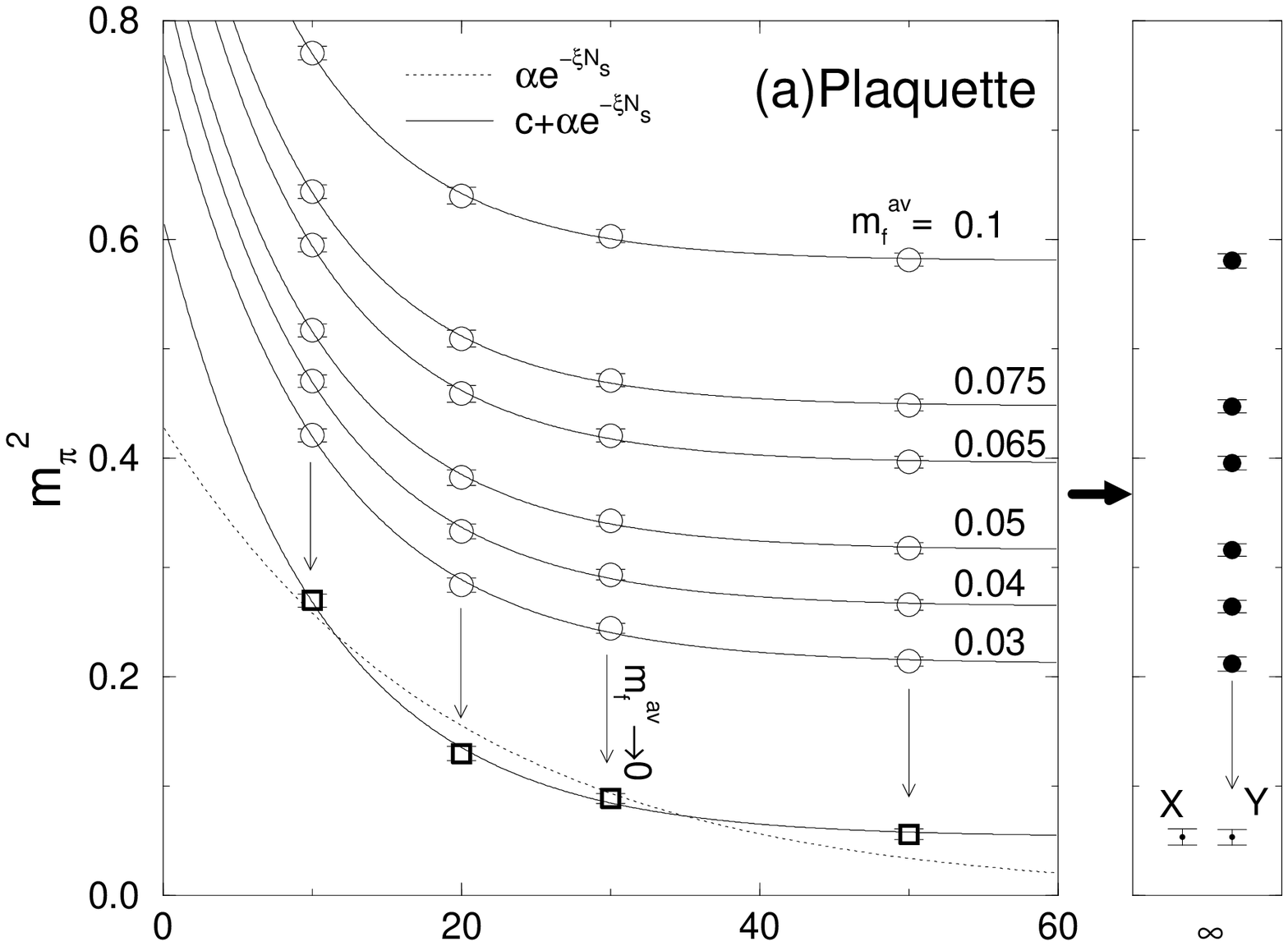}
  \vspace{-4pt}
  \epsfxsize=7.6cm \epsfbox{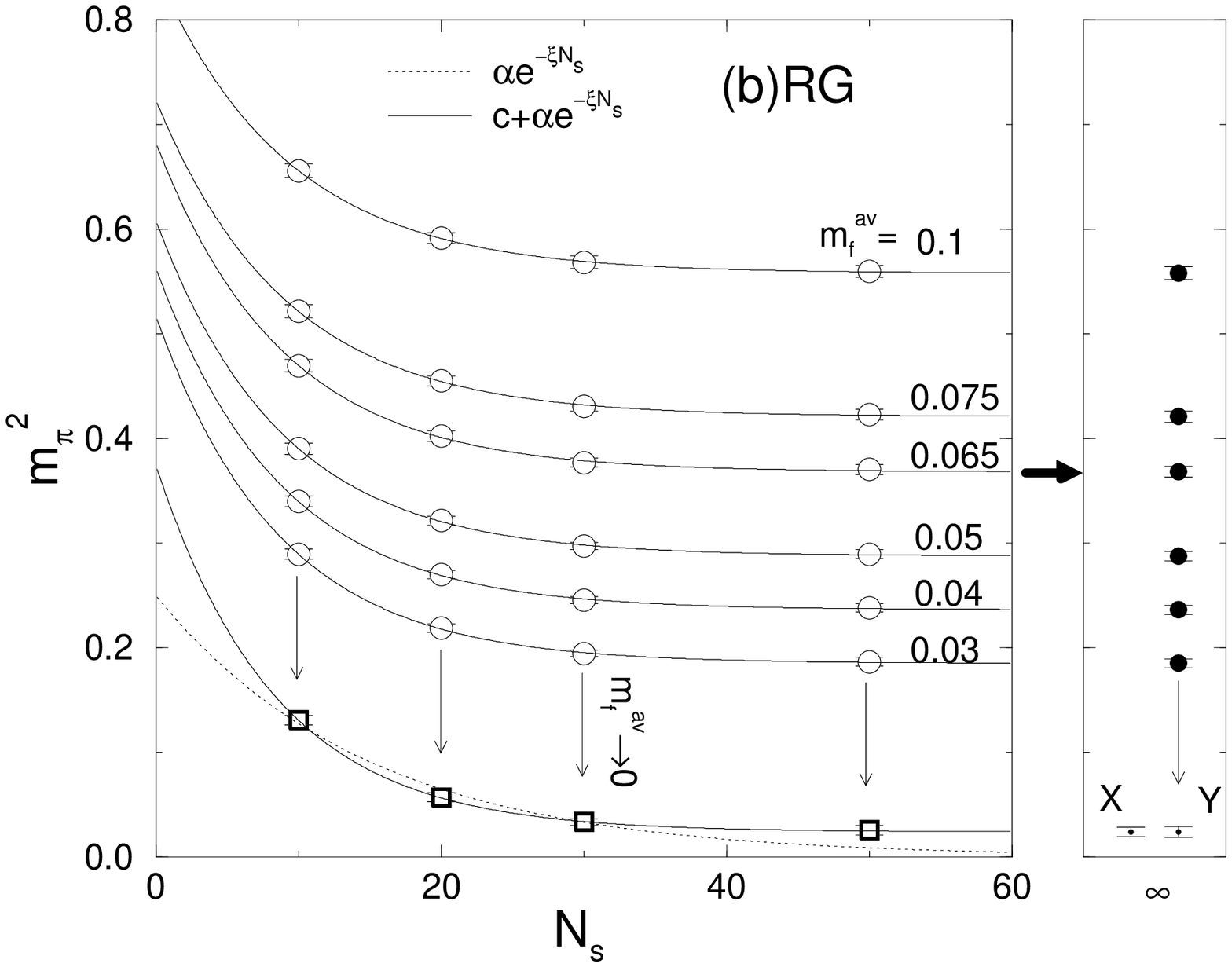}
  \vspace{-24pt}
  \caption{Pion mass squared as a function of $N_s$ at $M=1.7$
    for the plaquette (a) and RG (b) actions.
    Two limits $m_f^{av}\rightarrow 0$ and $N_s\rightarrow \infty$, and 
    their combinations in two possible orders denoted by ``X'' and ``Y'' 
    are shown.}
  \label{fig:pi2-Ns}
  \vspace{-20pt}
\end{figure}

In Fig.~\ref{fig:pi2-mfNs} the pion mass squared $m_\pi^2$ is plotted
as a function of the averaged bare quark mass $m_f^{av}$ at $M=1.7$ 
in the case of the plaquette gauge action.
Since the linearity of $m_\pi^2$ in $m_f^{av}$ is well satisfied, 
we adopt a linear chiral extrapolation in $m_f^{av}$.  As it can be seen in
the Fig.~\ref{fig:pi2-mfNs}, we find a non-zero value for $m_\pi^2$ at $m_f^{av} = 0$ 
for each $N_s$.

This non-zero value, however, is expected to vanish exponentially 
as $N_s \rightarrow \infty$. 
To see whether this occurs, 
we plot $m_\pi^2$ at $m_f^{av}=0$ by thick square symbols 
in Fig.~\ref{fig:pi2-Ns}(a) as a function of $N_s$. 
Fitting these points by the form $\alpha e^{-\xi N_s}$
yields the dotted line with  $\chi^2/dof=21.4$.  
An alternative fit, allowing a constant, $ c + \alpha e^{-\xi N_s}$
gives the solid line with $\chi^2/dof=1.7$ and $ c= 0.0532(75)$. 
Clearly the latter fit better reproduces the behavior of our data. 
A similar phenomenon has been previously reported also
at $\beta=5.7$~\cite{Columbia1}.

In order to confirm the existence of a non-zero $c$, 
we attempt to interchange the order of the limits $m_f^{av}\rightarrow 0$ and
$N_s\rightarrow\infty$.  As shown by solid lines going through open circles 
in  Fig.~\ref{fig:pi2-Ns}(a), we first make a fit of form
$m_\pi^2 (m_f^{av},N_s) = c^\prime(m_f^{av}) + \alpha e^{-\xi N_s}$ 
to make an $N_s\rightarrow\infty$ extrapolation for each value of $m_f^{av}$. 
The results, shown by solid circles in Fig.~\ref{fig:pi2-mfNs}, 
exhibits a linear behavior in $m_f^{av}$.  A linear chiral extrapolation 
$m_\pi^2(m_f^{av},N_s=\infty) = c^\prime(m_f^{av}) 
= d + \gamma m_f^{av}$ then yields 
$d = 0.0531(70)$, 
which agrees well with the value of $c$ previously obtained.
The commutativity of the two limits is summarized in Fig.~\ref{fig:pi2-Ns}(a), 
where the two values obtained for $m_\pi^2(m_f^{av}=0, N_s=\infty)$ are marked 
by ``X'' and ``Y''.

The above analyses strongly support the conclusion that
$m_\pi^2$ at $m_f^{av} = 0$ does not vanish in the limit $ N_s =\infty$
at $a^{-1} \simeq $ 1 GeV for the 4-dimensional lattice size of $12^3\times 24$.
We find this conclusion to apply as well to the RG-improved gauge action, as 
shown in Fig.~\ref{fig:pi2-Ns}(b). 

\begin{figure}[t]
  \vspace{-16pt}
  \begin{center}
    \leavevmode
  \epsfxsize=6.4cm \epsfbox{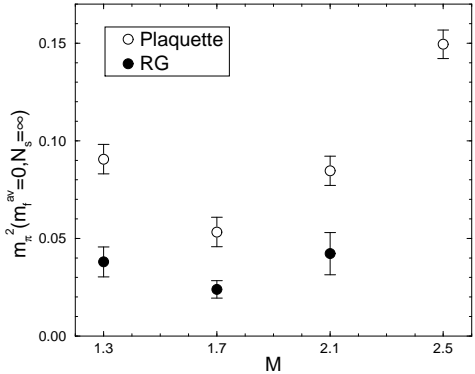}
\end{center}
\vspace{-32pt}
  \caption{Summary of the squared pion mass in the double limit
    $m_f\rightarrow 0$ and $N_s\rightarrow\infty$.}
  \label{fig:pi2-M}
  \vspace{-16pt}
\end{figure}

In Fig.~\ref{fig:pi2-M}
the residual $m_\pi^2(m_f^{av} = 0, N_s =\infty)$ is plotted as a function 
of $M$ for the plaquette and RG-improved gauge actions.
We observe that a non-vanishing residual remains under the 
variation of $M$ for both 
actions.  An improvement may be seen for the RG action, however, 
in that the magnitude of the residual 
is reduced by about a factor of two compared to that for the plaquette action. 

\begin{figure}[t]
  \vspace{-6pt}
  \epsfxsize=7cm \epsfbox{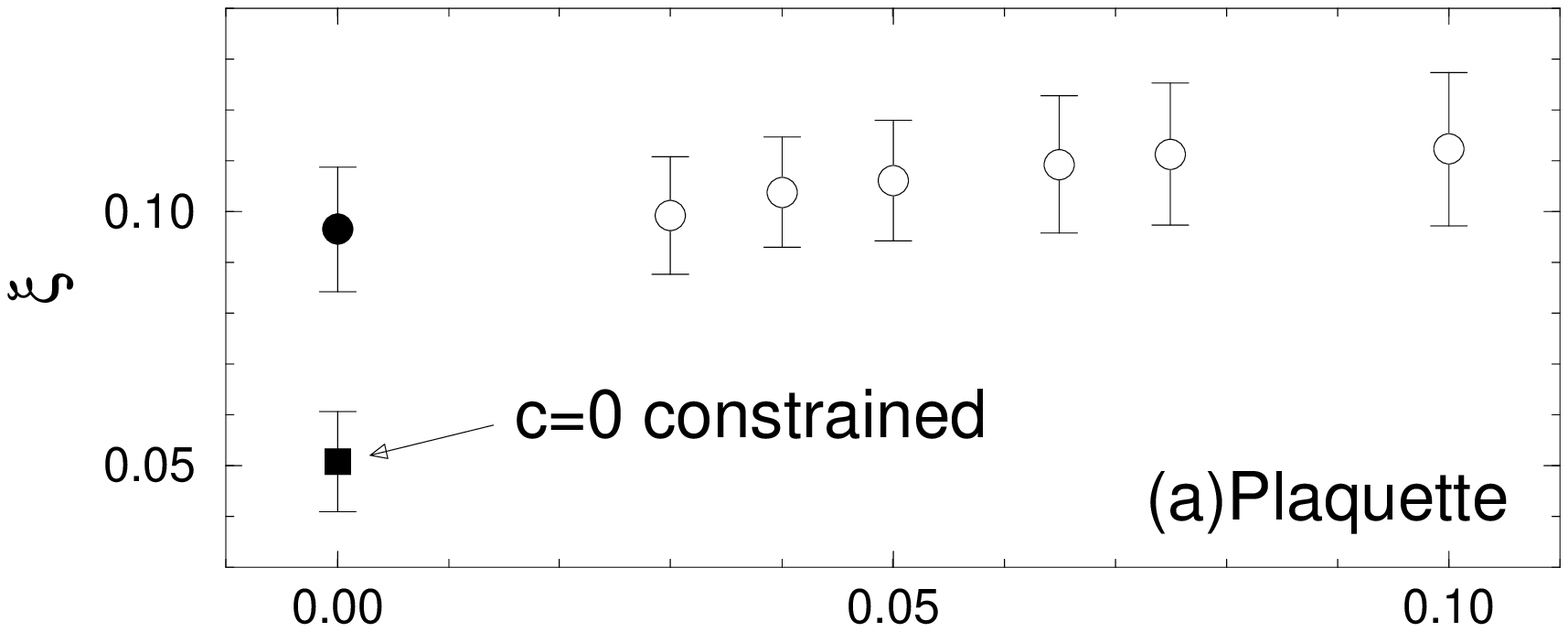}
  \vspace{-4pt}
  \epsfxsize=7cm \epsfbox{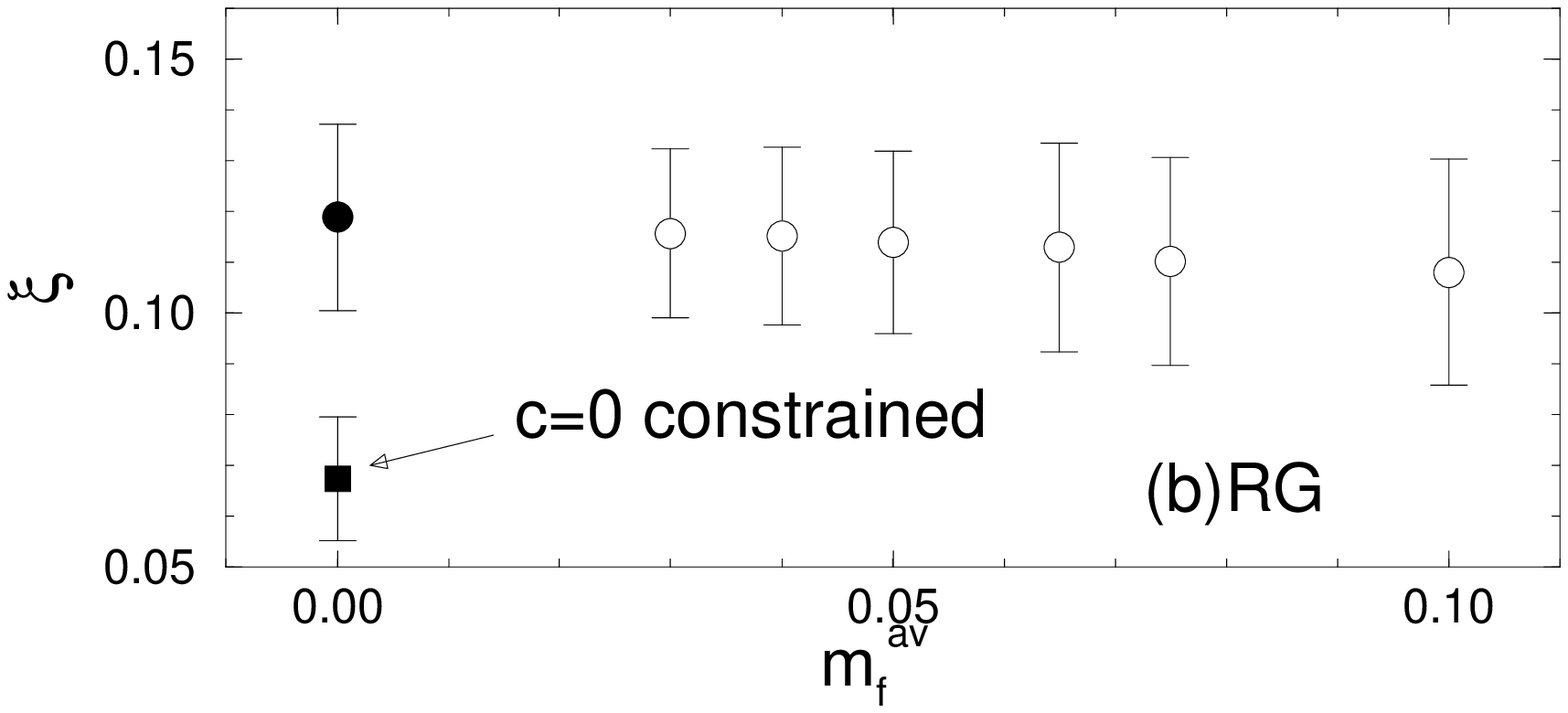}
  \vspace{-24pt}
  \caption{Decay rate $\xi$ from the curves of Figs.~\protect\ref{fig:pi2-Ns}
for the plaquette (a) and RG (b).}
  \label{fig:dlate-mf}
  \vspace{-16pt}
\end{figure}

Finally the decay rate $\xi$ extracted from the fits is shown 
in Fig.\ref{fig:dlate-mf} as a function of $m_f^{av}$ at $M=1.7$. 
For both actions, the chiral extrapolation of $\xi$ from $m_f^{av}\ne 0$ (open circles) 
smoothly agrees with the one directly obtained at 
$m_f^{av}=0$ assuming $c\ne 0$ (filled circle), but disagrees with
the value from the $c=0$ fit.
Furthermore the values of $\xi$ varies little with $m_f^{av}$. This is 
consistent with the expectation that the decay rate is governed by the transfer 
matrix in the extra dimension,  which is independent of $m_f$.

\section{DISCUSSION}

Our study of domain-wall QCD with the plaquette gauge action has shown that 
the pion mass in the chiral limit remains non-zero for 
an infinitely sized extra dimension 
at a strong gauge coupling corresponding to $a^{-1} \simeq$ 1 GeV 
on a $12^3\times 24$ lattice. We have also found that this conclusion remains 
unchanged for the RG-improved gauge action.

One possibility for the origin of a non-zero pion mass is 
finite spatial size effects, which is being checked by increasing the 
spatial size from 12 to 16. 
Another possibility is that it is an artifact of the 
linear chiral extrapolation, 
which does not take into account the possible presence of quenched 
chiral logarithms.  To explore this possibility,
the chiral breaking term in the axial Ward-Takahashi identity~\cite{FS}, 
which we expect to be free from the quenched singularities, 
is currently being investigated.

Finally it is possible that no chiral zero modes exist at $a^{-1} \simeq$ 1 GeV, 
and therefore the domain-wall formalism fails to realize chiral symmetry 
in the region of strong coupling. 
A plausible explanation for this failure might be that the range of
$M$ for zero modes to exist disappears for stronger couplings, where
there is no gap of eigenvalues for the 4-dimensional Wilson operator
\cite{eigen}.

\vspace*{12pt}

This work is supported in part by the Grants-in-Aid
of Ministry of Education
(Nos. 09304029, 10640246, 10640248, 11640250, 10740107, 11640294, 11740162). 
SE, TI, KN and YT are JSPS Research Fellows.
AAK and TM are supported by the Research for the Future 
Program of JSPS.


\begin{thebibliography}{9}
\bibitem{Kaplan} D.~Kaplan, Phys.\ Lett.\ B288 (1992) 342.
\bibitem{FS} V.~Furman and Y.~Shamir, Nucl.\ Phys.\ B439 (1995) 54.
\bibitem{BlumSoni} T.~Blum and A.~Soni, Phys.\ Rev.\ D56 (1997) 174;
  Phys.\ Rev.\ Lett.\ 79 (1997) 3595.
\bibitem{BlumRev} {\it For a review, see}
  T.~Blum, Nucl.\ Phys.\ B (Proc. Suppl.) 73 (1999) 167.
\bibitem{RG} Univ.\ of Tsukuba report UTHEP-118 (1983), unpublished.
\bibitem{Columbia1} P.~Chen {\it et al.}, 
  Nucl.\ Phys.\ B (Proc. Suppl.) 73 (1999) 204.
\bibitem{eigen}R.~Setoodeh, C.~T.~H.~Davies and I.~M.~Barbour,
Phys.\ Lett.\ B213 (1988) 195.
\end{thebibliography}
\end{document}